\documentclass[envcountsame]{llncs}

\usepackage{color}
\usepackage[dvips,breaklinks, raiselinks, 
             colorlinks,bookmarks,bookmarksnumbered,
             citecolor={black},linkcolor={black},filecolor={black},
             pagecolor={black},urlcolor={black},
             pdfpagemode={None}, pdfstartview={Fit},
             pdfstartpage={1}, pdftitle={ Automatic Verification of Parametric
               Specifications with Complex Topologies},
             pdfauthor={ Johannes Faber and Carsten Ihlemann and Swen Jacobs and
               Viorica Sofronie-Stokkermans },
             pdfsubject={ }]
            {hyperref} 

\usepackage[dvips]{graphicx}
\graphicspath{{./figures/}}
\usepackage{pstricks}
\usepackage{rotating}
\usepackage{amsmath}
\usepackage{amssymb}

\usepackage{oz,ozjh,csp-oz,dc,ozjf}
\usepackage[notheorems]{jf}

\usepackage{subfigure}
\usepackage{wrapfig}
\usepackage{tikz}
\usetikzlibrary{arrows}
\usetikzlibrary{backgrounds}

\makeatletter
\newcommand{\codlabel}[1]{(\stepcounter{equation}\immediate\write\@auxout{%
\string\newlabel{#1}{{\theequation}{}{}{}{}}}\theequation)}
\makeatother
\def\dcchop{\mathrel{\cat}}

\newcommand{\hpilot}{H-PILoT\xspace}
\newcommand{\syspect}{Syspect\xspace}
\renewcommand{\cod}{COD\xspace}

\newcommand{\bbR}{\mathbb{R}}

\newcommand{\mT}{\mathcal{T}}
\newcommand{\mK}{\mathcal{K}}

\newcommand{\upperLine}[1]{\setbox0=\hbox{#1}%
\setbox1=\hbox spread -1\wd0{\hspace{11cm}}%
#1\raisebox{0.5ex}{\rule{\wd1}{.5pt}}}
\newcommand{\lowerLine}{\raisebox{0.5ex}{\rule{11cm}{.5pt}}}

\makeatletter
\providecommand*{\toclevel@author}{1}
\providecommand*{\toclevel@title}{1}
\makeatother

\title{Automatic Verification of Parametric Specifications with
Complex Topologies\thanks{\avacssupportshorter}}

\author{Johannes Faber\inst{1}
\and Carsten Ihlemann\inst{2} 
\and Swen Jacobs\inst{3}
\and\\Viorica Sofronie-Stokkermans\inst{2}
}

\institute{Department of Computing Science, University of Oldenburg, Germany
\and
Max-Planck-Institut f{\"u}r Informatik, Saarbr{\"u}cken, Germany
\and
\'Ecole Polytechnique F\'ed\'erale de Lausanne, Switzerland\\
}

\showCorrections

\begin{document}

\maketitle

\pagestyle{headings}

\begin{abstract}
The focus of this paper is on reducing the complexity in verification 
by exploiting modularity at various levels:  in specification, 
in verification, and structurally.
For specifications, we use the modular language CSP-OZ-DC, which allows us to 
decouple verification tasks concerning data from those concerning durations.
At the verification level, we exploit modularity in theorem proving
for rich data structures and use this
for invariant checking.
At the structural level, we analyze possibilities for modular verification of
systems consisting of various components which interact.
We illustrate these ideas by automatically verifying safety properties
of a case study from the European Train Control System standard, which extends
previous examples by comprising a complex track topology with lists of
track segments and trains with different routes.
\end{abstract}



\section{Introduction}
\label{sec:introduction}
Parametric real-time systems
arise in a natural way in a wide range of applications, including controllers 
for systems of cars, trains, and planes. Since many such
systems are safety-critical, there is great interest in methods for
ensuring that they are safe. 
In order to verify such systems, one needs (i) suitable formalizations 
and (ii) efficient verification techniques.
In this paper, we analyze both aspects. 
Our main focus throughout the 
paper will be on reducing complexity by exploiting modularity at 
various levels: in the specification, in verification, and also 
structurally. 
The main contributions of the paper are: 
\begin{itemize}
\item[(1)] We exploit modularity at the specification 
level. In Sect.~\ref{Sec:ModularSpecifications}, we use 
the modular language CSP-OZ-DC (COD), 
which allows us to separately specify processes (as Communicating Sequential 
Processes, CSP), data (using Object-Z, OZ) and time (using the Duration
Calculus, DC). 

\item[(2)] We exploit modularity in verification (Sect.~\ref{Sec:Verification}). 

\begin{itemize}
\item First, we consider transition constraint systems (TCSs) that can be automatically obtained from the COD
specification, and address verification tasks such as invariant checking.
We show that for pointer data structures, we can obtain decision procedures for these verification tasks.

\item Then we analyze situations 
in which the use of COD specifications allows us to decouple 
verification tasks concerning data (OZ) from verification tasks 
concerning durations (DC). 
For systems with a parametric number of components, this allows us to impose
 (and verify) conditions on the single components which guarantee safety of the
overall complex system.
\end{itemize}
 
\item[(3)] We also use modularity at a structural
level. In Sect.~\ref{Sec:ComplexTopologies}, we use results from 
\cite{Sofronie-getco06} to obtain possibilities for
modular verification of systems with complex topologies by decomposing
them into subsystems with simpler topologies.

\item[(4)] We describe a tool chain which translates a graphical UML version of the CSP-OZ-DC specification into TCSs, and automatically verifies the specification using our prover \hpilot and other existing tools (Sect.~\ref{sec:tools}). 

\item[(5)] We illustrate the ideas on a running example taken from the European Train Control System standard
(a system  with a complex topology and a parametric number 
of components---modeled using pointer data 
structures and parametric constraints), 
and present a way of fully automatizing verification (for given safety 
invariants) using our tool chain.  
\end{itemize}

\noindent {\bf Related work.}
Model-based development and verification of railway 
control systems with a complex track topology are analyzed in 
\cite{HP2007}. The systems are described in a domain-specific language and
translated into SystemC code that is verified using bounded model
checking. Neither verification of systems with a parametric number
of components nor pointer structures are examined there.

In existing work on the verification of 
parametric systems often only
few aspects of parametricity are studied together. 
\cite{PQ2009} addresses the verification of 
temporal properties for hybrid systems (in particular also fragments of the
ETCS as case study) but only supports parametricity in the data domain. 
\cite{AJ1998} presents 
a method for the verification of a parametric number of timed automata
with real-valued clocks, while in \cite{APRXZ01} only finite-state processes are
considered. 
In \cite{AJN+2004}, regular model checking for a parametric number 
of homogeneous linear processes and systems operating on queues or stacks is
presented.
There is also work on the analysis of
safety properties for parametrized systems with an arbitrary number of
processes operating on unbounded integer variables
\cite{ADR2009,CTV06,LahiBrya04}. In contrast to ours, these methods sacrifice
completeness by using either an over-approximation of the transition relation or
abstractions of the state space. We, on the other hand, 
offer complete methods (based on decision procedures for data structures) 
for problems such as invariant checking and bounded model checking.

\begin{figure}[t]
  \centering 
  \vspace{-4ex}
  \begin{minipage}[t]{5.6cm}
    \includegraphics[width=5.4cm]{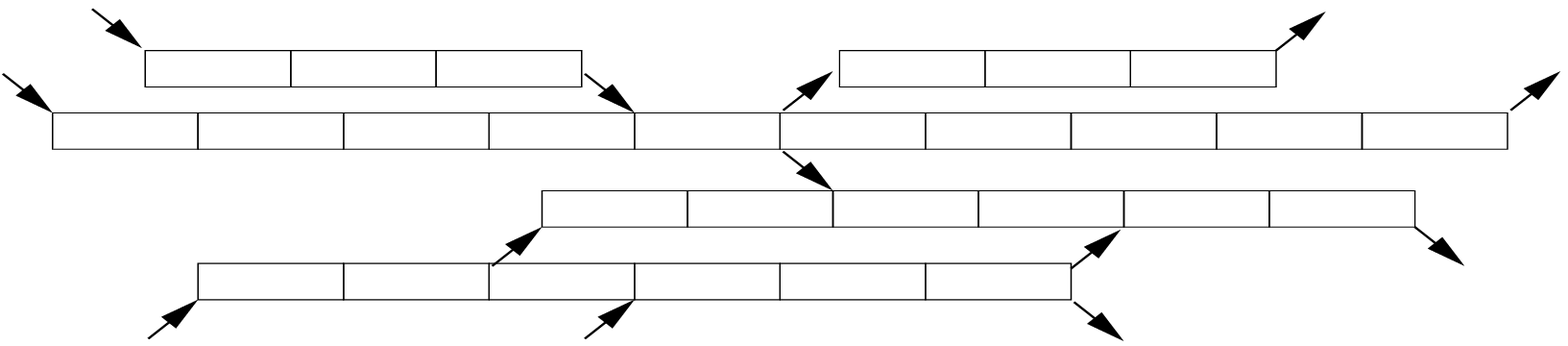}
    \caption{Complex Track Topology}
    \label{Fig:TrackTopology}
  \end{minipage}
  \hfil
  \begin{minipage}[t]{5cm}
    \includegraphics[width=5cm]{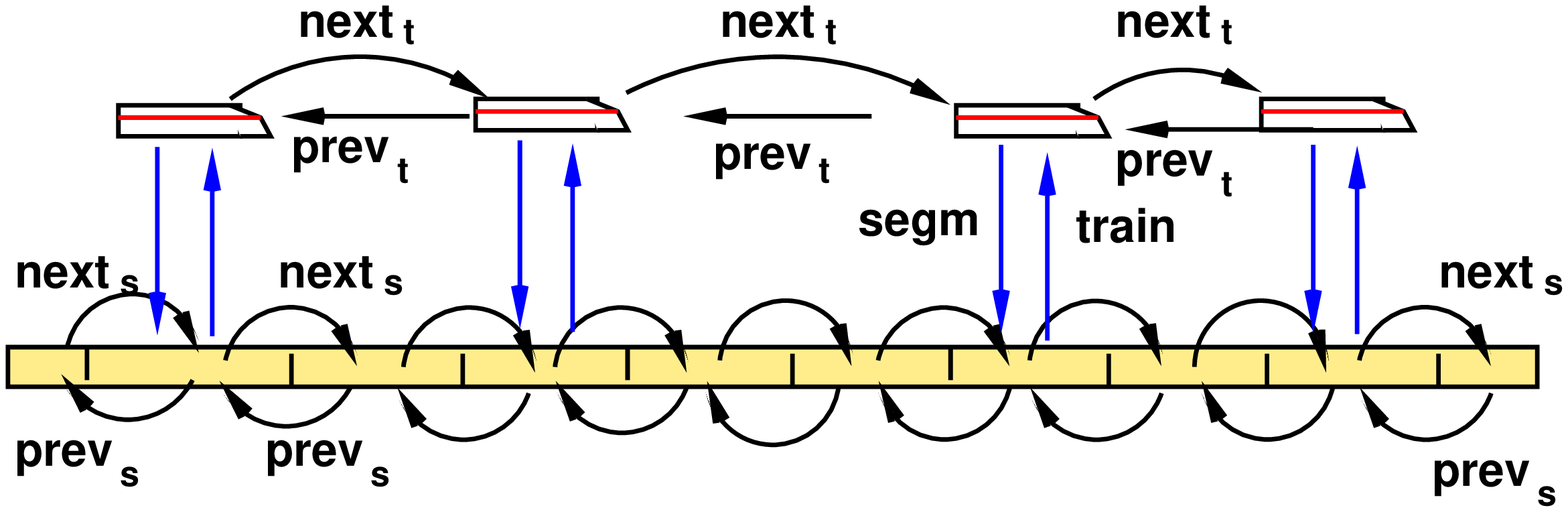}
    \caption{Linear Track Topology}
    \vspace{-7mm} 
    \label{Fig:simple-track}
  \end{minipage}
  \vspace{-3ex}
\end{figure}

\medskip
\noindent 
{\bf Motivating example.}
\label{sec:illustration}
Consider a system of trains on a complex track topology 
as depicted 
in Fig.~\ref{Fig:TrackTopology}, and  
a radio block center (RBC) that has information about track segments and trains, like e.g. length, occupying train and allowed maximal speed for segments, and current position, segment and speed for trains.
We will show under which situations safety of the 
system with complex track topology is a consequence of safety of systems 
with linear track topology. 
Such modular verification possibilities allow us to 
consider the verification of a simplified version of this example, 
consisting of a 
{\em linear track} (representing a concrete route in the track topology), 
on which trains are 
allowed to enter or leave at given points. 
We model a general RBC controller for an  
area with a linear track topology and an arbitrary number of trains.
For this, we use
a theory of pointers with
sorts ${\sf t}$ (for trains; {\sf next}$_t$ returns the next train on the track)
and ${\sf s}$ (for segments; with {\sf next}$_s$, {\sf prev}$_s$ describing the
next/previous segment on the linear track). The link between trains and  
segments is described by appropriate functions ${\sf train}$ and ${\sf segm}$
(cf.\ Fig.~\ref{Fig:simple-track}). 
%
In addition, we integrated a simple timed train controller ${\sf Train}$ 
into the model. This allowed us to certify
      that certain preconditions for the verification of the RBC are
      met by every train which  satisfies the specification 
of ${\sf Train}$, 
by reasoning on the timed and the untimed part of the system independently.
%

\section{Modular Specifications: CSP-OZ-DC}
\label{Sec:ModularSpecifications}

We start by presenting the specification language CSP-OZ-DC (COD)
\cite{HO02b,Hoenicke2006} which allows us to present in a modular 
way the control flow, data changes, and timing aspects
of the systems we want to verify. 
We use \emph{Communicating Sequential Processes} (CSP) to specify the
control flow of a system using processes
over events; \emph{Object-Z} (OZ) for describing the state space and its 
change, and the \emph{Duration
  Calculus} (DC) for modeling (dense) real-time
constraints over durations of events. 
The operational semantics of COD is defined in \cite{Hoenicke2006} in terms of a
timed 
automata model. 
For details on CSP-OZ-DC and its semantics, we refer to 
\cite{HO02b,Hoenicke2006,faber-jacobs-sofronie-07}. 
Our benefits from using \cod are twofold:
\begin{itemize}
\item \cod is
compositional in the sense that it suffices to prove safety properties for the
separate components to prove safety of the entire system \cite{Hoenicke2006}. 
This makes it possible to use different verification techniques for different
parts of the specification, e.g. for control structure and timing properties.
\item We benefit from high-level tool support given by \syspect\footnote{\url{http://csd.informatik.uni-oldenburg.de/~syspect/}}, a UML editor for a
dedicated UML profile \cite{MOR+2008} proposed to formally model real-time
systems. It has a semantics in terms of COD. 
Thus, \syspect serves as an easy-to-use
front-end to formal real-time specifications, with a graphical user interface.
\end{itemize}
\subsection{Example:  Systems of Trains on Linear Tracks}
\label{sect:csp-oz-dc-trains} 
To illustrate the ideas, we present some aspects of the case study 
mentioned in Sect.\ \ref{sec:illustration} (the full case study 
is presented in \cite{atr66}). 
We exploit the benefits of COD in (i) the specification of  
a complex RBC controller; (ii) the specification of a 
controller for individual trains; and (iii) composing such specifications.
Even though space does not allow us to present all details, 
we present aspects of the example which cannot be considered with 
other formalisms, and show how to cope in a natural way with parametricity.

\medskip
\noindent
{\bf CSP part.} The processes and their interdependency is specified using the CSP
specification language. The RBC system passes
repeatedly through four phases, modeled by events with corresponding \cod
schemata $updSpd$ (\emph{speed update}), $req$ (\emph{request update}), 
$alloc$ (\emph{allocation update}), and $updPos$ (\emph{position update}).  

{\scriptsize \upperLine{\em CSP: }
}

\scalebox{.65}{
$\begin{array}{@{}l}
        \method\ enter: [ s1? : Segment ; t0? : Train ; t1? : Train ; t2? :
        Train ]\\[0.3ex] 
        \method\ leave: [ ls? : Segment ; lt? : Train ]\\[0.3ex]
        \lchan\ alloc, req, updPos, updSpd
\end{array}$
}

\smallskip

\scalebox{.65}{
$\begin{array}{@{}l@{}l@{}l@{}l@{}l@{}l@{}l@{}l@{}l@{}l@{}l@{}l@{}}
\main & \stackrel{c}{=} & ((updSpd {\rightarrow} State1)~~~ &  State1 & \stackrel{c}{=} & ((req {\rightarrow} State2)~~~ &             State2 & \stackrel{c}{=} & ((alloc {\rightarrow} State3)~~~ &  State3 & \stackrel{c}{=} & ((updPos {\rightarrow} \main) \\[0.3ex]
          & \Box & (leave {\rightarrow} \main) &  &   \Box & (leave {\rightarrow} State1) &  &             \Box & (leave {\rightarrow} State2) & &             \Box & (leave {\rightarrow} State3) \\[0.3ex]
          & \Box & (enter {\rightarrow} \main)) & &            \Box & (enter {\rightarrow} State1))& &             \Box & (enter {\rightarrow} State2)) & &             \Box & (enter {\rightarrow} State3))
\end{array}$
}

\lowerLine

\noindent 
The \emph{speed update} models the fact that every train chooses its 
speed according to its 
knowledge about itself and its track segment as well as the next
track segment. The \emph{request update} models how trains send a request for
permission to enter the next segment when they come close to the end of their
current segment. 
The \emph{allocation update} models how the RBC may either
grant these requests by allocating track segments to trains that have made a
request, or allocate segments to trains that are not currently on the route and
want to enter. The \emph{position update} models how all trains report their
current positions to the RBC, which in turn 
de-allocates segments that have been left and gives movement authorities to the
trains.
Between any of these four updates, we can have trains \emph{leaving} or
\emph{entering} the track at specific segments using the events $leave$ and
$enter$. The effects of these updates are defined in the OZ part. 

\medskip
\noindent
{\bf OZ part. } The OZ part of the specification consists of data classes,
axioms, the \Init schema, and update rules.

\smallskip
\noindent
{\bf Data classes.}
The data classes declare function symbols that can change 
their values during runs of the
system, and are used in the OZ
part of the specification.

\scalebox{.65}{
\begin{minipage}{1.5\linewidth}
\begin{sidebyside}
      \begin{schema}{SegmentData}\\
          train : Segment \fun Train 
          \comment{Train on segment}\\
          req : Segment \fun \num 
          \comment{Requested by train}          \\
          alloc : Segment \fun \num
          \comment{Allocated by train}          \\
      \end{schema}
      \nextside
      \begin{schema}{TrainData}\\
          segm : Train \fun Segment 
          \comment{Train segment}\\
          next : Train \fun Train 
          \comment{Next train}\\
          spd : Train \fun \real 
          \comment{Speed}\\
          pos : Train \fun \real
          \comment{Current position} \\
          prev : Train \fun Train
          \comment{Prev.~train}
      \end{schema}
    \end{sidebyside}
\end{minipage}
}

\noindent
{\bf Axioms.}
The axiomatic part defines properties of the data structures and system
parameters which do not change during an execution of the system:
$gmax: \bbR$ (the global maximum speed),
$decmax: \bbR$ (the maximum deceleration of trains),
$d: \bbR$ (a safety distance between trains),
and $bd: \bbR \rightarrow \bbR$ (mapping the speed of a train to a safe
approximation of the corresponding braking distance).
We specify properties of those parameters, among which 
an important one is $d \geq bd ( gmax ) + gmax \cdot \Delta t$  
stating that the safety distance $d$ to the end of the segment is greater than
the braking distance of a train at maximal speed $gmax$ plus a further 
safety margin (distance for driving $\Delta t$ time units at speed $gmax$).
Furthermore, unique, non-negative ids for trains (sort
$Train$) and track segments (sort $Segment$) are defined. The route is
modeled as a doubly-linked 
list\footnote{Note that 
  we use 
  relatively loose axiomatizations of 
  the list structures for both trains and segments, also allowing for
  disjoint families of linear, possibly infinite lists.}
 of track segments, where every
segment has additional properties specified by the constraints in the state
schema.  

\begin{wrapfigure}[8]{r}{.5\linewidth}
  \noindent \scalebox{.65}{
    \noindent $\begin{array}{@{}l} 
~~\\[-15ex]
\hline
      ~~\\[-2ex] 
      \forall t : Train @ tid ( t ) > 0 \\[0.3ex]
      \forall t1 , t2 : Train | t1 \neq t2 @ tid ( t1 ) \neq tid ( t2 ) \\
      \forall s : Segment @ prevs ( nexts ( s ) ) = s \\
      \forall s : Segment @ nexts ( prevs ( s ) ) = s \\
      \forall s : Segment @ sid ( s ) > 0 \\
      \forall s : Segment @ sid ( nexts ( s ) ) > sid ( s ) \\
      \forall s1 , s2 : Segment | s1 \neq s2 @ sid ( s1 ) \neq sid ( s2 ) \\
      \forall s : Segment | s \neq snil @ length ( s ) > d + gmax \cdot \Delta t \\
      \forall s : Segment | s \neq snil @ 0 < lmax ( s ) \land lmax ( s ) \leq gmax \\
      \forall s : Segment @ lmax ( s ) \geq lmax ( prevs ( s ) ) - decmax  \cdot \Delta t\\
      \forall s1 , s2 : Segment @ tid ( incoming ( s1 ) ) \neq tid ( train ( s2 ) )~\text{(*)}\\[1.4ex]
      \hline
    \end{array}$
  }
\end{wrapfigure}

E.g., $sid$ is increasing along the $nexts$ pointer, the $length$ of a
segment is bounded from below in 
terms of $d$ and $gmax \cdot \Delta t$, and the difference
between local maximal speeds on neighboring segments is bounded by $decmax \cdot \Delta t$.
Finally, we have a function $incoming$,
the value of which is either a train which wants to enter the given segment from
outside the current route, or $tnil$ if there is no such 
train. Although the valuation of $incoming$ can change during an execution, we
consider the constraint (*) as a property of 
our environment that always holds. 
Apart from that, incoming may change
arbitrarily and is not explicitly updated.
Note that $Train$ and $Segment$
are pointer sorts with a special null element ($tnil$ and $snil$, respectively),
and all constraints implicitly only hold for non-null elements.
So, constraint (*) actually means
\vspace{-3mm}
{\small 
\begin{multline*}
  \forall s1 , s2 : Segment | s1 \neq snil \neq s2 \land incoming(s1) \neq tnil
  \land train(s2) \neq tnil \\
  @ tid ( incoming ( s1 ) ) \neq tid ( train ( s2 ) )\\[-4ex]
\end{multline*}
}

\begin{wrapfigure}[7]{r}{.5\linewidth}
  \vspace{-7ex}
  \hspace{-3.5ex}
  \scalebox{.65}{
    \begin{minipage}{1.5\linewidth}
      \begin{schema}{\Init}
        \forall t : Train @ train ( segm ( t ) ) = t \\
        \forall t : Train @ next ( prev ( t ) ) = t \\
        \forall t : Train @ prev ( next ( t ) ) = t \\
        \forall t : Train @ 0 \leq pos ( t ) \leq length ( segm ( t ) ) \\
        \forall t : Train @ 0 \leq spd ( t ) \leq lmax ( segm ( t ) ) \\
        \forall t : Train @ alloc ( segm ( t ) ) = tid ( t )\\
        \forall t : Train @ alloc ( nexts ( segm ( t ) ) ) = tid ( t ) \\
        \t1 \lor length ( segm ( t ) ) - bd ( spd ( t ) ) > pos ( t ) \\
        \forall s : Segment @ segm ( train ( s ) ) = s
      \end{schema}
    \end{minipage}
  }
\end{wrapfigure}
 
\vspace{-4mm}
\noindent {\bf Init schema.} 
The \emph{{\sf Init} schema}  describes the initial state of the
system.  It essentially states that trains are arranged 
in a doubly-linked list,   
that all trains are initially placed correctly on the track segments and that
all trains respect their speed limits.

\medskip
\noindent 
{\bf Update rules.}
Updates of the state space, that are executed when the corresponding event from
the CSP part is performed, are specified with \emph{effect
  schemata}. 
The schema for $updSpd$, for instance, consists of three rules, 
distinguishing (i) trains whose distance to the end of the segment 
is greater than the safety distance $d$ (the first two lines 
of the constraint), (ii) trains that are beyond 
the safety distance near the end of the segment, and for which the next segment is
allocated, and (iii) trains that are near the end of the segment without an 
allocation. In
case (i), the train can choose an arbitrary speed below the maximal speed
of the current segment. In case (ii), the train needs to brake if the
speed limit of the next segment is below the current limit. In case (iii),
the train needs to brake such that it safely stops before reaching the end of 
the segment.

\vspace{-2mm}

{\scriptsize
\noindent \begin{effectop}{updSpd}
\Delta ( spd ) \\
\where
\forall t : Train | pos ( t ) < length ( segm ( t ) ) - d \land spd ( t ) -
decmax  \cdot \Delta t > 0 \\
\quad @ \max\{0, spd ( t ) - decmax \cdot \Delta t\} \leq spd' ( t )  \leq  lmax ( segm (
t ) ) \\
\forall t : Train | pos ( t ) \geq length ( segm ( t ) ) - d \land alloc (
nexts ( segm ( t ) ) ) = tid ( t ) \\
\quad@ 
\max\{0, spd ( t ) - decmax \cdot \Delta t\} \leq spd' ( t ) \leq \min\{lmax ( segm ( t ) ),
lmax ( nexts ( segm ( t ) ) )\} \\ 
\forall t : Train | pos ( t ) \geq length ( segm ( t ) ) - d \land \lnot alloc (
nexts ( segm ( t ) ) ) = tid ( t ) \\
\quad@ spd' ( t ) =  \max\{0, spd ( t ) - decmax \cdot \Delta t \}
\end{effectop}
}

\vspace{-3mm}

\noindent {\bf Timed train controller.}\\
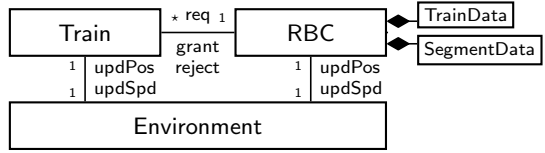
\begin{wrapfigure}[6]{r}{7.2cm}
{\small  \vspace{-1.7cm}
    \begin{tikzpicture}[node distance=2cm]
    \sf
    \tikzstyle{sdt}=[rectangle,draw=black,fill=white,thick,inner sep=2mm]
    \tikzstyle{solidline}=[>=latex,semithick,->, auto]
    \tikzstyle{strb}=[sdt,minimum width=2cm,node distance=3cm]

    \path
    node[strb,label={[yshift=2mm]right:\tiny $\star$}]       (trn) {Train}
    node[strb,right of=trn,label={[yshift=2mm]left:\tiny 1}] (rbc) {RBC}
    node[sdt,minimum width=5cm, node distance=1.2cm] (env) 
                             at ([xshift=1.5cm,yshift=-1.2cm]trn.mid |- rbc.mid) 
                                                                   {Environment}
    ;

    \path[solidline,-]
    (trn) edge node[anchor=north] 
                  { \parbox{1cm}{\centering\scriptsize grant\\reject}}
               node[anchor=south] 
                  { \parbox{1cm}{\centering\scriptsize req}}(rbc)
    (rbc) edge                    (rbc)
    (trn) edge node { \parbox{1cm}{\scriptsize updPos\\updSpd}} 
                       node[anchor=north east] {\tiny 1}
                       node[anchor=south east] {\tiny 1}      (trn |- env.north)
    (rbc) edge node { \parbox{1cm}{\scriptsize updPos\\updSpd}} 
                       node[anchor=north east] {\tiny 1}
                       node[anchor=south east] {\tiny 1}      (rbc |- env.north)
    ;

    \path[anchor=west]
    node[sdt,inner sep=.8mm]   (tdt) at ([xshift=4mm,yshift=-.9mm]rbc.north east)
                                                          {\scriptsize TrainData}
    node[sdt,inner sep=.8mm]   (sdt) at ([xshift=4mm,yshift=.9mm]rbc.south east)
                                                       {\scriptsize SegmentData};

    \path[solidline,diamond-]
        ([yshift=1.5mm] rbc.east) edge ([yshift=1.5mm] tdt.west  |- rbc.east)
        ([yshift=-1.5mm]rbc.east) edge ([yshift=-1.5mm]sdt.west  |- rbc.east);

  \end{tikzpicture}
}  \vspace{-4.7ex}
  \caption{Structural overview}
  \label{fig:classes}

\end{wrapfigure}

\vspace{-8mm}
\noindent
In the DC part of a specification, real-time constraints are specified:
A second, timed controller {\sf Train} 
(for one train only) interacts with the RBC controller, which is 
presented in the overview of the case study in Fig.~\ref{fig:classes}. 
The train controller {\sf Train} 
consists of three timed components running in parallel. The first 
updates the train's position.  
This component contains e.g.\ the DC formula
\[
 \neg (true\dcchop \dcevent updPos \dcchop (\ell < \Delta t)\dcchop \dcevent updPos \dcchop true),
\]
that specifies a lower time bound $\Delta t$ on $updPos$ events.
The second component checks periodically
whether the train is beyond the safety distance to the end of the segment. 
Then, it starts braking within a short reaction time. 
The third component requests an extension of the movement authority from the RBC,
which may be granted or rejected.
The full train controller can be found in
\cite{atr66}.

\section{Modular Verification}
\label{Sec:Verification}
In this section, we combine two approaches for the
verification of safety properties of COD specifications: 
\begin{itemize}
\item We introduce
the invariant checking approach  
and present decidability results for local theory extensions 
that imply decidability of the invariant checking problem for a large
class of parameterized systems. 
\item We illustrate 
how we can combine this invariant checking for the RBC specification 
with a method for model checking of real-time properties (introduced in \cite{MFH+2008}) for the 
COD specification for a single train {\sf Train}.
\end{itemize}
 Formally, our approach works on a \emph{transition constraint
system} (TCS) obtained from the COD specification by an automatic translation
(see \cite{faber-jacobs-sofronie-07}) which is guaranteed to capture the defined
semantics of COD (as defined in \cite{Hoenicke2006}).
\begin{definition}
The tuple $T = (V, \Sigma, {\sf (Init)}, {\sf (Update)})$ is a \emph{transition
  constraint system},  
 which specifies:
the variables ($V$) and function symbols ($\Sigma$) 
whose values may change over time; 
a formula ${\sf (Init)}$ specifying the properties of initial states; 
and a formula ${\sf (Update)}$
which specifies the transition relation in terms of the values 
of variables $x \in V$ and function symbols $f \in \Sigma$ before a transition
and their values (denoted $x'$, $f'$) after the transition. 
\end{definition}
In addition to the TCS, we obtain a {\em background theory} ${\cal T}$ from the
specification, describing properties of the used data structures and system
parameters that do not change over time. Typically, $\mT$ consists of 
a family of standard theories (like the theory of real numbers), 
axiomatizations for data structures, and constraints on system
parameters.
In what follows $\phi {\models_{\cal T}} \psi$ denotes 
logical entailment and means that every model of the theory ${\cal T}$ 
which is a model of $\phi$ is also a model for $\psi$. 
We denote {\sf false} by $\perp$, so $\phi {\models_{\cal T}} {\perp}$ 
means that $\phi$ is unsatisfiable w.r.t. ${\cal T}$.

\subsection{Verification Problems}
\label{sec:invar-check-param}
We consider the problem of 
{\em invariant checking}
of safety properties.\footnote{We can address bounded model checking problems in
  a similar way,
  cf.\ \cite{jacobs-sofronie07,faber-jacobs-sofronie-07,Sofronie-Ihlemann-Jacobs-tacas08}.}
To show that a safety property, represented as a formula ${\sf (Safe)}$,
is an invariant of a TCS $T$
(for a given background theory ${\cal T}$), we need
to identify an \emph{inductive invariant} ${\sf (Inv)}$ which strengthens ${\sf
  (Safe)}$, i.e., we need to prove that 
\begin{itemize}
\item[(1)] ${\sf (Inv) \models_\mT (Safe)}$,
\item[(2)] ${\sf (Init) \models_\mT (Inv)}$, and 
\item[(3)] ${\sf (Inv) \wedge (Update) \models_\mT (Inv')}$,
where ${\sf (Inv')}$ results from ${\sf (Inv)}$ by replacing each $x \in
V$ by $x'$ and each $f \in \Sigma$ by $f'.$
\end{itemize}
\begin{lemma}
If  ${\sf (Inv)}, {\sf (Init)}$ and 
${\sf (Update)}$ belong to a class of formulae for which the entailment 
problems w.r.t. ${\cal T}$ above are decidable 
then the problem of checking 
that ${\sf (Inv)}$ is an invariant of $T$ 
(resp.\ $T$ satisfies the property 
${\sf (Safe)}$) is decidable. 
\end{lemma}
We use this result in a verification-design loop as follows: 
We start from a specification written in COD. We use a translation 
to TCS and check whether a certain formula $({\sf Inv})$ 
(usually a safety property) is an inductive invariant.

\noindent (i) If invariance can be proved, safety of the system is guaranteed.

\noindent (ii) If invariance cannot be proved, we have the following possibilities: 
\begin{itemize}
\item[1.] Use a specialized prover to construct a 
counterexample (model in which the property $({\sf Inv})$ is not an invariant) 
which can be used to find errors in the specification and/or to strengthen 
the invariant\footnote{This last step is the only part which is not fully automatized. For future work we plan to investigate possibilities of
automated invariant generation or strengthening.}.
\item[2.] Using results in 
\cite{sofronie-ijcar10} we can often derive 
additional (weakest) constraints on the parameters which guarantee that 
${\sf Inv}$ is an invariant. 
\end{itemize}
Of course, the decidability results for the theories used in the 
description of a system can be also used for 
checking consistency of the specification. 

If a TCS models a system with a parametric number of 
components, the formulae in problems (1)--(3) may contain 
universal quantifiers (to describe properties of all components), hence  
standard SMT methods -- which are only complete
for ground formulae -- 
do not yield decision procedures.
In particular, for (ii)(1,2) and for consistency checks 
we need possibilities of reliably 
detecting satisfiability of sets of universally quantified formulae 
for which standard SMT solvers cannot be used. 
We now present situations in which this is possible.

\subsection{Modularity in Automated Reasoning: Decision Procedures}
\label{Sec:HierarchicalReasoning}
We identify classes of theories for which 
invariant checking (and bounded model checking) is decidable. 
Let ${\cal T}_0$ be a theory with signature $\Pi = (S_0, \Sigma_0, {\sf Pred})$,
where $S_0$ is a set of sorts, and $\Sigma_0$ and ${\sf Pred}$ are sets of 
function resp.\ predicate symbols. We consider extensions of 
${\cal T}_0$ with new function symbols in a set
$\Sigma$, whose properties are axiomatized by a set ${\cal K}$ of clauses.  

\medskip
\noindent {\bf Local theory extensions.} We are interested in 
theory extensions in which 
for every set $G$ of ground clauses we can effectively determine a finite 
(preferably small) set of instances of the axioms 
${\cal K}$ sufficient for checking satisfiability of $G$ without 
loss of completeness. 
If $G$ is a set of $\Pi^c$-clauses (where 
$\Pi^c$ is the extension of $\Pi$ with constants in
a set $\Sigma_c$), we denote by ${\sf st}({\cal K}, G)$
the set of ground terms starting with a $\Sigma$-function symbol 
occurring in ${\cal K}$ or $G$, 
and by ${\cal K}[G]$ the
set of instances of ${\cal K}$ in which the terms starting with
$\Sigma$-functions are in ${\sf st}({\cal K}, G)$.
${\cal T}_0 {\cup} {\cal K}$ is a {\em local extension} of 
${\cal T}_0$ 
\cite{Sofronie-cade-05} if the following condition holds:

\medskip
\noindent \begin{tabular}{ll}
~~~{\sf (Loc)} & For every set $G$ of ground clauses, 
$G \models_{{\cal T}_0 {\cup} {\cal K}} \perp$ 
 iff ${\cal K}[G] \cup G \models_{{\cal T}_0^{\Sigma}} \perp$ 
\end{tabular}

\medskip
\noindent where ${{\cal T}_0^{\Sigma}}$ is the extension of 
${\cal T}_0$ with the free functions in $\Sigma$. 
We can define {\em stable locality} ({\sf SLoc}) 
in which we use the set ${\cal K}^{[G]}$ of instances of 
$\mK$ in which the 
variables below $\Sigma$-functions are instantiated with terms in 
${\sf st}({\cal K},G)$.
In local theory extensions, sound and complete 
hierarchical reasoning is possible.
\begin{theorem}[\cite{Sofronie-cade-05}]
With the notations introduced 
above, if ${\cal T}_0 \subseteq {\cal T}_0 \cup {\cal K}$ satisfies
condition $({\sf (S)Loc})$ then the following are equivalent to 
$G \models_{{\cal T}_0 {\cup} {\cal K}} \perp$: 
\begin{itemize}
\item[(1)] ${\cal K}{*}[G] {\cup} G \models_{{\cal T}_0^{\Sigma}} \perp$ (${\cal K}{*}[G]$ is ${\cal K}[G]$ for local; ${\cal K}^{[G]}$ for stably local extensions).
\item[(2)] ${\cal K}_0 {\cup} G_0 {\cup} D \models_{{\cal T}_0^{\Sigma}} \perp$, 
where ${\cal K}_0 {\cup} G_0 \cup D$ is obtained from ${\cal K}{*}[G] {\cup} G$
by introducing (bottom-up) new  constants $c_t$ for subterms 
$t = f(g_1, \dots, g_n)$ with $f \in \Sigma$, $g_i$ 
ground $\Sigma_0 \cup \Sigma_c$-terms; replacing the terms with the corresponding constants; and adding 
the definitions $c_t \approx t$ to the set $D$. 
\item[(3)] ${\cal K}_0 {\cup} G_0 {\cup} N_0 \models_{{\cal T}_0} \perp$, where\\
$\displaystyle{~~~ N_0  = \{ \bigwedge_{i = 1}^n c_i \approx d_i \rightarrow c = d \mid 
f(c_1, \dots, c_n) \approx c, f(d_1, \dots, d_n)\approx d \in D \}}.$ \\[-2ex]

\end{itemize}
\label{lemma-rel-transl}
\end{theorem}
The hierarchical reduction method is implemented in the system 
\hpilot 
\cite{Sofronie-Ihlemann-hpilot08}. 
\begin{corollary}[\cite{Sofronie-cade-05}]
\label{compl} 
If the theory extension 
${\cal T}_0 \subseteq {\cal T}_1$ satisfies  
${\sf ((S)Loc)}$  
then satisfiability of sets of ground clauses  
$G$ w.r.t.\ ${\cal T}_1$ is decidable 
if ${\cal K}{*}[G]$ is finite and 
${\cal K}_0 {\cup} G_0 {\cup} N_0$ belongs to 
a decidable fragment ${\cal F}$ of ${\cal T}_0$. 
Since the size of  
${\cal K}_0 {\cup} G_0 {\cup} N_0$ is polynomial in the size of $G$
(for a given ${\cal K}$), locality allows us to express the complexity 
of the ground satisfiability problem w.r.t.\ ${\cal T}_1$  
as a function of the complexity of the satisfiability 
of ${\cal F}$-formulae w.r.t.\ ${\cal T}_0$.
\end{corollary}
\subsection{Examples of Local Theory Extensions}
\label{Sect:local}
We are interested in reasoning efficiently about 
data structures and about updates of data structures. 
We here give examples of such theories.

\medskip
\noindent {\bf Update axioms.} 
In \cite{Sofronie-Ihlemann-Jacobs-tacas08} we show that 
update rules  ${\sf Update}(\Sigma, \Sigma')$ which 
describe how the values of the $\Sigma$-functions change,  
depending on a set $\{ \phi_i \mid i \in I \}$ of 
mutually exclusive conditions, 
define local theory extensions. 
\begin{theorem}[\cite{Sofronie-Ihlemann-Jacobs-tacas08}]
\label{updates}
Assume that $\{ \phi_i \mid i \in I \}$
are formulae over the base signature
such that 
$\phi_i({\overline x}) \wedge \phi_j({\overline x}) \models_{{\cal T}_0} \bot 
\text{ for } i {\neq} j$, and that 
$s_i, t_i$ are (possibly equal) terms over the signature $\Sigma$ such that ${\cal T}_0
\models \forall {\overline x} (\phi_i({\overline x}) {\rightarrow} s_i({\overline x}) {\leq} t_i({\overline x}))$ for all $i \in I$. 
Then the extension of ${\cal T}_0$ with axioms of the form ${\sf Def}(f)$ is local.

\smallskip
${\sf Def}(f)~~~~~ \forall {\overline x} (\phi_i({\overline x}) \rightarrow s_i({\overline x}) \leq f({\overline x}) \leq t_i({\overline x}))  i \in I.$

\label{thm:updates}
\end{theorem}
{\bf Data structures.}
Numerous locality results for data structures exist, e.g. 
for fragments of the theories of arrays 
\cite{manna-vmcai-06,Sofronie-Ihlemann-Jacobs-tacas08}, 
and pointers 
\cite{NeculaMcPeak,Sofronie-Ihlemann-Jacobs-tacas08}. 
As an illustration -- since the model we used 
in the running example involves a theory 
of linked data structures -- we now 
present a slight extension of the fragment of the theory of pointers  
studied in \cite{NeculaMcPeak,Sofronie-Ihlemann-Jacobs-tacas08}, 
which is useful for modeling the track topologies and 
successions of trains on these tracks. 
We consider a set of pointer sorts ${\sf P} =
\{ {\sf p}_1,\ldots,{\sf p}_n \}$ and a scalar sort ${\sf  s}$.\footnote{We 
assume that we only have one scalar sort for simplicity of 
presentation; the scalar theory can 
itself be an extension or combination of theories.} 
Let $(\Sigma_s, {\sf Pred}_s)$ be a scalar 
signature, and let $\Sigma_P$ be a set of  function symbols with arguments 
of pointer sort consisting of sets 
$\Sigma_{{\overline p} \rightarrow {\sf s}}$ (the family of 
functions of arity ${\overline p} {\rightarrow} {\sf s}$), 
and $\Sigma_{{\overline p} \rightarrow {\sf p}}$ (the family of 
functions of arity ${\overline p} {\rightarrow} {\sf p}_i$). (Here ${\overline p}$ is a tuple ${\sf p}_{i_1} \dots {\sf p}_{i_k}$ with $k \geq 0$.)
We assume that for every pointer sort ${\sf p} \in {\sf P}$, 
$\Sigma_P$ contains a constant ${\sf null}_{\sf p}$ of sort ${\sf p}$. 
\begin{example}
The fact that we also allow scalar fields with 
more than one argument is very useful because it allows, for instance, to model 
certain relationships between different nodes. Examples of such scalar
fields could be: 
\begin{itemize}
\item ${\sf distance}(p, q)$ associates with non-null $p, q$ of pointer type 
a real number;
\item  ${\sf reachable}(p, q)$ associates with non-null $p, q$ of pointer type 
a boolean value (true (1) if $q$ is reachable from $p$ using the next functions, 
false (0) otherwise). 
\end{itemize}
\end{example}
Let $\Sigma = \Sigma_P \cup \Sigma_s$.
In addition to allowing several pointer types and functions of 
arbitrary arity, we loosen some of the 
restrictions imposed in 
\cite{NeculaMcPeak,Sofronie-Ihlemann-Jacobs-tacas08}. 
\begin{definition}
An \emph{extended pointer clause} 
is a formula of form
$\forall \bar{p}. ~~ ({\cal E} \vee \varphi)(\bar{p})$, 
where $\bar{p}$
is a set of pointer variables including all 
free variables of ${\cal E}$ and $\varphi$, and: 
\begin{itemize}
\item[(1)] ${\cal E}$ consists of  disjunctions of pointer equalities,
and has the property that for every term $t = f(t_1, \dots, t_k)$ 
with $f \in \Sigma_P$ occurring in ${\cal E} \vee \varphi$, 
${\cal E}$ contains an atom of the form $t' = {\sf null}_{\sf p}$ 
for every proper subterm (of sort ${\sf p}$) $t'$ of $t$;
\item[(2)] $\varphi$ is an \emph{arbitrary}
formula of sort ${\sf s}$.
\end{itemize}
${\cal E}$ and $\varphi$ may additionally 
contain free scalar and pointer constants, and $\varphi$ may contain additional 
quantified variables of sort ${\sf s}$.
\label{def:pointers}
\end{definition}
\begin{theorem}
Let $\Sigma = \Sigma_P \cup \Sigma_s$ be 
a signature as defined before. Let 
${\cal T}_s$ be a theory of scalars with signature $\Sigma_s$. 
Let $\Phi$ be a set of $\Sigma$-extended pointer clauses. 
Then, for every set $G$ of ground 
clauses over an extension $\Sigma^c$ of $\Sigma$ with constants in a countable 
set $c$ the following are equivalent:
\begin{itemize}
\item[(1)] $G$ is unsatisfiable w.r.t.\ $\Phi \cup {\cal T}_s$; 
\item[(2)]  $\Phi^{[G]} \cup G$ is an unsatisfiable set of clauses 
in the disjoint combination ${\cal T}_s \cup {\cal E}{\cal Q}_P$ of 
${\cal T}_s$ and ${\cal E}{\cal Q}_P$, the 
many-sorted theory of pure equality over pointer sorts,  
\end{itemize}
where $\Phi^{[G]}$ consists of all instances of $\Phi$ in which 
the universally quantified variables of pointer type occurring in $\Phi$
are replaced by ground terms of pointer type in 
the set ${\sf st}(\Phi, G)$ 
of all ground terms of sort $p$ occurring in $\Phi$ or in $G$.
\label{thm:pointers}
\end{theorem}
The proof is similar to that in \cite{Sofronie-Ihlemann-Jacobs-tacas08}. 
\hpilot can be used as a decision 
procedure for this theory of pointers 
-- if the theory of scalars 
is decidable -- and for any extension of this theory with function updates
in  the fragment in Thm.~\ref{updates}.  
\begin{example} 
Let $P = \{ {\sf sg}(segment), {\sf t}(train) \}$, and let ${\sf next_t}, {\sf prev_t} : {\sf t}
\rightarrow {\sf t}$, and ${\sf next_s},  
{\sf prev_s} : {\sf sg} \rightarrow {\sf sg}$, and 
${\sf train} : {\sf sg} \rightarrow {\sf t}$, 
${\sf segm} : {\sf t} \rightarrow {\sf sg}$, and 
functions of scalar sort as listed at the beginning of 
Sect.~\ref{sect:csp-oz-dc-trains}.
All axioms describing the background theory and the
initial state in Sect.~\ref{sect:csp-oz-dc-trains} 
are expressed by extended pointer clauses. 

\smallskip
\noindent
The following formula expressing a property of reachability of trains 
can be expressed as a pointer clause: 

\smallskip
\noindent 
{\small $\forall p, q (p \neq {\sf null}_t \wedge q \neq {\sf null}_t \wedge 
{\sf next}_t(q) \neq {\sf null}_t \rightarrow ({\sf reachable}(p,q) \rightarrow 
{\sf reachable}(p,{\sf next}_t(q))).$}
\end{example}

\noindent {\bf Decidability for verification.}
A direct consequence of Thm.~\ref{lemma-rel-transl} and Cor.~\ref{compl} 
is the following decidability result for invariant checking: 
\begin{corollary}[\cite{sofronie-ijcar10}]
Let $T$ be the transition constraint system and ${\cal T}$ be the background theory associated with a
specification. 
If the update rules ${\sf Update}$ and the invariant property ${\sf Inv}$ 
can be expressed as sets of 
clauses which define a chain of local theory 
extensions ${\cal T} \subseteq {\cal T} \cup {\sf Inv}({\overline x},{\overline f}) \subseteq  {\cal T} \cup {\sf Inv}({\overline x},{\overline f}) \cup {\sf Update}({\overline x}, {\overline x'}, {\overline f}, {\overline f'})$ 
then checking whether a formula is an invariant is decidable. 
\label{cor:inv-ch-gen}
\end{corollary}
In this case we can use \hpilot as a decision procedure (and also 
to construct a model in which the property ${\sf Inv}$ is not an invariant).
We can also use results from \cite{sofronie-ijcar10} to 
derive additional (weakest) constraints on the parameters which guarantee 
that ${\sf Inv}$ is an invariant.

\subsection{Example: Verification of the Case Study} 
We demonstrate how the example from Sect.~\ref{sec:illustration} can be verified by
using a combination of the invariant checking approach presented in
Sect.~\ref{sec:invar-check-param} and a model checking  approach for timing properties. This
combination is necessary because the example contains both the RBC component
with its discrete updates, and the train controller {\sf Train} 
with real-time safety properties. Among other things, the specification of the RBC assumes that the
train controllers always react in time to make the train brake before reaching a critical position. 

Using the modularity of COD, we can separately use the invariant checking approach to verify the RBC for a parametric number of trains, and the approach for model checking  DC formulae to verify that
every train satisfies the timing assumptions made in the RBC specification.

\medskip
\noindent
{\bf Verification of the RBC.}
The verification problems for the RBC are 
satisfiability problems containing universally quantified formulae, hence 
cannot be decided by standard methods of reasoning in combinations of 
theories. Instead, we use the hierarchical reasoning
approach from Sect.~\ref{Sec:HierarchicalReasoning}. 

\medskip
\noindent {\em Safety properties.} 
As safety property for the RBC we want to prove that we never have two trains on
the same segment:
\[{\sf (Safe)} := \forall t_1, t_2: Train.\ t_1 \neq t_2 \rightarrow 
id_s(segm(t_1)) \neq id_s(segm(t_2)).\] 
To this end, we need to find a formula ${\sf (Inv)}$ such that we can prove
{\small \begin{itemize}
\item[(1)] ${\sf (Inv)} \cup \neg {\sf (Safe)} \models_\mT \bot$,
\item[(2)] ${\sf (Init)} \cup \neg {\sf (Inv)} \models_\mT \bot$, and
\item[(3)] ${\sf (Inv) \cup {\sf (Update)} \cup \neg (Inv')} \models_\mT \bot$, 
\end{itemize}}

\noindent 
where ${\sf (Update)}$ is the update formula associated with the 
transition relation obtained by translating the COD
specification  into TCS
\cite{Hoenicke2006,faber-jacobs-sofronie-07}, and ${\sf (Init)}$ consists of the
constraints in the $\Init$ schema. The background theory $\mT$ is
obtained from the state schema of the OZ part of the specification: it is the
combination of the theories of real numbers and integers,
together with function and constant symbols satisfying the constraints given in
the state schema.

Calling H-PILoT on problem (3) with {\sf (Inv)} = {\sf (Safe)} 
shows us that {\sf (Safe)}
is not inductive over all transitions.
Since we expect the updates to preserve the well-formedness
 properties in {\sf (Init)}, we tried to use this as our invariant,
but with the same result.
However, inspection of counterexamples provided by \hpilot
allowed us to identify
the following additional constraints
needed to make the invariant inductive:
{\small 
\begin{align*}
  {\sf (Ind_1)} := \forall t: Train.\ & pc \neq InitState \land alloc(nexts(segm(t)))
  \neq tid(t)\\  
  & \rightarrow length(segm(t)) - bd(spd(t)) > pos(t) +
  spd(t) \cdot \Delta t\\
  {\sf (Ind_2)} := \forall t: Train.\ & pc \neq InitState \land pos(t)
  \geq length(segm(t)) - d\\ 
  & \rightarrow spd(t) \leq lmax(nexts(segm(t)))\label{eq:1}
\end{align*}%
}%
The program counter $pc$ is introduced in the translation process from COD to
TCS and we use the constraint $pc \neq InitState$ to indicate that the system is
not in its initial location.
Thus, define ${\sf (Inv)}$ as the conjunction $({\sf Init}) \land ({\sf Ind_1}) \land ({\sf Ind_2})$.
Now, all of the verification tasks above can automatically be proved using
\syspect and \hpilot, in case (3) after splitting the problem into a number of
sub-problems. 
To ensure
that our system is not trivially safe because of inconsistent assumptions, we
also check for consistency of $\mT$, ${\sf (Inv)}$ and ${\sf (Update)}$.
Since by Thm.~\ref{thm:updates}
all the update rules in the RBC specification define local theory extensions,
and the axioms specifying properties of the data types 
are extended pointer clauses, by Cor.~\ref{cor:inv-ch-gen} we obtain the following decidability result.
\begin{corollary}
Checking properties (1)--(3) is decidable for all formulae ${\sf Inv}$
expressed as sets of extended pointer clauses with the property 
that the scalar part belongs to a decidable fragment of the theory 
of scalars.
\end{corollary}
{\em Topological invariants.} We also considered  
certain topological invariants of the system -- e.g.\ that 
if a train $t$ is inserted between trains $t_1$ and $t_2$, 
the {\sf next} and {\sf prev} links are adjusted properly, and if a 
train leaves a track then its ${\sf next}_t$ and ${\sf prev}_t$ links become 
${\sf null}$.  We also checked that if certain reachability 
conditions -- modeled using a binary transitive function ${\sf reachable}$ 
with Boolean output which is updated when trains enter or 
leave the line track -- are satisfied before an insertion/removal of trains
then they are satisfied also after. 
We cannot include these examples in detail here; 
they will be presented in a separate paper.

\medskip
\noindent
{\bf Verification of the timed train controller.}
Using the model checking approach from \cite{MFH+2008}, we can automatically
prove real-time properties of COD specifications. 
In this case, we use the
approach only on the train controller part {\sf Train} (Fig.~\ref{fig:classes}).
We show that the safety
distance $d$ and the braking distance $bd$ postulated in the RBC controller model can actually be achieved by trains that comply with the train specification. That is, we prove that (for an arbitrary train) the train position
$curPos$ is never beyond its movement authority $ma$:
\[{\sf (Safe_T)} := \neg\dceve( curPos > ma ).\]

\noindent
{\bf Safety of the overall system.}
The safety property for trains ${\sf (Safe_T)}$ implies that train controllers satisfying the specification also
satisfy the timing assumptions made implicitly in the RBC
controller. Compositionality of COD guarantees \cite{Hoenicke2006} that it is
sufficient to verify
these components separately. Thus, by additionally proving that ${\sf (Inv)}$ is a safety invariant of the RBC, we have shown that the system consisting of a combination of the RBC controller and arbitrarily many train controllers is safe.


\section{Modular Verification for  Complex Track Topologies}
\label{Sec:ComplexTopologies}
We now consider a complex track as described in Fig.~\ref{Fig:TrackTopology}.
Assume that the track 
can be modeled as a directed graph $G = (V, E)$ with the following properties: 
\begin{itemize}
\item[(i)] The graph $G$ is acyclic (the rail track does not contain cycles); 
\item[(ii)] The in-degree of every node is at most 2 (at every point at 
which two lines meet, at most two linear tracks are merged).
\end{itemize}
\begin{theorem}
For every track topology satisfying conditions 
(i) and (ii) above we can find a decomposition 
${\cal L} = \{ {\sf ltrack}_i \mid i \in I \}$ into linear tracks such that 
if $(x, y) \in E$ then $y = {\sf next}_s^{{\sf ltrack}_i}(x)$ for some $i \in I$ 
and for every ${\sf ltrack} \in {\cal L}$ identifiers are increasing w.r.t.\ 
${\sf next}^{\sf ltrack}_s$. 
\label{thm:acyclic}
\end{theorem}
We assume that for each linear track ${\sf ltrack}$
we have one controller $RBC^{\sf ltrack}$ which uses the control 
protocol described in Sect.~\ref{sect:csp-oz-dc-trains}, where   
we label the functions describing the 
train and segment succession using indices 
(e.g.\ we use ${\sf next}^{\sf ltrack}_t, {\sf prev}^{\sf ltrack}_t$ for 
the successor/predecessor of a train on {\sf ltrack}, 
and ${\sf next}^{\sf ltrack}_s$, ${\sf prev}^{\sf ltrack}_s$ for 
the successor/predecessor of a segment 
on {\sf ltrack}. 
Assume that these controllers 
are compatible on their common parts, i.e.\ 
(1) if two tracks ${\sf track_1}, {\sf track_2}$ 
have a common subtrack ${\sf track_3}$ then the corresponding fields 
agree, i.e.\ whenever $s, {\sf next}^{\sf track_i}_s(s)$ are on ${\sf track}_3$, ${\sf next}^{\sf track_1}_s(s) {=} {\sf next}^{\sf track_2}_s(s) {=} {\sf next}^{\sf track_3}_s(s)$ (the same for  ${\sf prev}_s$, and
for ${\sf next}_t, {\sf prev}_t$ on the corresponding tracks); (2) the update rules are compatible for 
trains jointly controlled.\footnote{We also assume 
that all priorities of the trains on the complex track are different.}
Under these conditions, proving safety for 
the complex track  can be reduced to checking safety of 
linear train tracks with incoming and outgoing trains 
(for details cf. \cite{atr66}).  
\begin{lemma}
A state $s$ of the system is a model 
$(P_{\sf t}, P_{\sf s}, {\mathbb R}, {\mathbb Z}, \{ {\sf next}^{\sf ltrack}, {\sf prev}^{\sf ltrack},$\\ ${\sf next}_s^{\sf ltrack}, {\sf prev}_s^{\sf ltrack} \}_{{\sf track} \in {\cal L}} \cup \{ {\sf segm}, {\sf train}, {\sf pos}, ... \})$, 
where all the functions relativized to tracks are compatible 
on common subtracks.
The following hold:
\begin{itemize}
\item[(a)] Every state $s$ of the 
system of trains on the complex track restricts to a state 
$s_{\sf ltrack}$ 
of the system of trains on its component linear track. 
\item[(b)] Any family $\{ s_{{\sf ltrack}_i} \mid i \in I \}$ 
of states on the component tracks which agree on the 
common sub-tracks can be ``glued together'' to a state $s$ 
of the system of trains on the complex track topology. 
\end{itemize}
(a) and (b) also hold if we consider {\em initial} states (i.e.\ states 
satisfying the initial conditions) and {\em safe} states (i.e.\ states 
satisfying the safety conditions in the invariant ${\sf Inv}$).
Similar properties hold for parallel actions and for transitions.\end{lemma}
\begin{theorem}
Consider a complex track topology satisfying conditions 
(i)--(ii) above. Let ${\cal L} = \{ {\sf ltrack}_i \mid i \in I \}$ be 
its decomposition 
into a finite  family of finite linear tracks 
such that for all 
 ${\sf ltrack}_1, {\sf ltrack}_2 \in {\cal L}$,  
${\cal L}$ contains all their common maximal linear subtracks. 
Assume that the tracks ${\sf ltrack}_i \in {\cal L}$ 
(with increasing segment identifiers w.r.t.\ ${\sf next}^{\sf ltrack}_s$) 
are controlled by controllers $RBC^{{\sf ltrack}_i}$ using  
the protocols in Sect.~\ref{sect:csp-oz-dc-trains} which synchronize on 
common subtracks. 
Then we can guarantee safety of the control protocol for the 
controller of the complex track obtained by interconnecting all 
linear track controllers $\{ RBC^{{\sf ltrack}_i} \mid i \in I \}$. 
\label{thm:modular}
\end{theorem}

\section{From Specification to Verification}
\label{sec:tools}

 \begin{figure}[t]
  \vspace{-3ex}
  \centering
  \begin{tikzpicture}[node distance=2cm]
    \sf
    \tikzstyle{tool}=[rectangle,rounded corners=5,
                      draw=black!50,fill=black!10,thick,inner sep=2mm]
    \tikzstyle{lang}=[rectangle,draw=black!50,fill=white,thick,inner sep=2mm,
                      dash pattern=on 1mm off 1mm]
    \tikzstyle{solidline}=[>=latex,semithick,->, auto]

    \path[anchor=west]            node [lang] (uml) {UML}              
    (uml.east)++(right:0.5cm)     node [lang] (cod) {CSP-OZ-DC}
    (cod.east)++(right:1.2cm)     node [tool] (pea) {PEA toolkit} 
    (pea.east)++(right:0.8cm)     node [tool] (hpt) {\hpilot}
    (hpt.west)++(up:0.7cm)       node [tool] (amc) {ARMC}
    (hpt.east)++(right:0.5cm)     node [tool] (prv) {Prover};
    
    \path[anchor=south]  
    (uml.north)++(up:1mm)         node        (sys) {\syspect};
    
    \path[solidline] 
             (uml)                    edge              (cod) 
             (cod)                    edge node {PEA}   (pea) 
             (pea)                    edge node[anchor=north] {TCS}   (hpt)
             (pea)                    edge node  {TCS}   (amc.west)
             ([yshift=1mm] hpt.east)  edge              ([yshift=1mm] prv.west)
             ([yshift=-1mm] prv.west) edge              ([yshift=-1mm] hpt.east);

    \begin{pgfonlayer}{background}
      \draw [tool] (sys.north -| sys.west) 
                   rectangle 
                   ([xshift=2mm,yshift=-2mm] cod.south -| cod.east);
    \end{pgfonlayer}
  \end{tikzpicture}
  \vspace{-3ex}
  \caption{Tool chain}
  \vspace{-4ex}
  \label{fig:toolchain}
\end{figure}
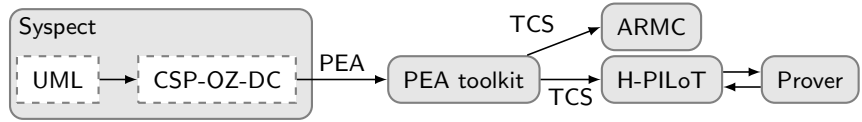

\noindent
For the practical application of verification techniques tool support is
essential. 
For this reason, in this section we introduce a full tool chain 
for automatically checking the invariance of 
safety properties starting from a given specification and give some experimental
results for our RBC case study. 

\medskip
\noindent {\bf Tool chain.}
\label{sec:translation}
The tool chain is sketched in Fig.~\ref{fig:toolchain}.
In order to capture the systems we want to verify, we use the COD front-end
\syspect (cf.~Sect.~\ref{Sec:ModularSpecifications}). 
\cite{Hoenicke2006} defines the semantics of \cod in terms of a
timed automata model called Phase Event Automata (PEA). 
A translation from PEA
into TCS is given in \cite{Hoenicke2006}, which is implemented in the PEA 
toolkit\footnote{\url{http://csd.informatik.uni-oldenburg.de/projects/epea.html}}
and used by \syspect. 

Given an invariance property, a \syspect model can directly be exported
into a TCS in the syntax of \hpilot. 
If the specification's background theory consists of chains of local theory
extensions, the user needs to specify via input dialog (i)~that the pointer
extension of \hpilot is to be used;
(ii)~which level of extension is used for each function symbol 
of the specification.
With this information, our tool chain can verify invariance of a safety
condition fully automatically by checking 
its invariance for 
each transition update (cf.~Sect.~\ref{sec:invar-check-param}). Therefore, for
each update, \syspect exports a file 
that is handed over to \hpilot. The safety invariance is proven 
if \hpilot detects the unsatisfiability of each verification task. 
Otherwise, \hpilot generates a model violating the invariance of the 
desired property, 
which may be used to fix the problems in the specification.

In addition, the PEA toolkit also supports output of TCS into the input language
of the abstraction refinement model checker ARMC \cite{PR2007}, which we used to
verify correctness of the timed train controller from our example.

\begin{wraptable}[8]{r}{4cm}
  \vspace{-5ex}
  \centering\scriptsize
  \begin{tabular}{lccc}
    \hline
                 & (sys) &  (hpi) & (yic) \\ 
    \hline
    ({\sf Inv}) \emph{unsat}   &    &  &   \\
     Part 1 & 11s     & 72s      & 52s \\ 
     Part 2 & 11s     & 124s     & 131s \\ 
    speed update & 11s     & 8s       & 45s \\ 
    \hline 
    {\sf(Safe)} \emph{sat}  & 9s     & 8s       & t.o.   \\ 
    \hline 
    Consistency & 13s     & 3s       & (U) 2s   \\ 
    \hline
  \end{tabular}\\
{(obtained on: AMD64, dual-core 2 GHz, 4 GB RAM)}
  
  \vspace{-3.5ex}
  \caption{Results}
  \label{tab:experiments}
\end{wraptable}

\medskip
\noindent {\bf Experimental results.}
\label{sec:experimental-results}
Table~\ref{tab:experiments} gives experimental results for checking the RBC
controller.\footnote{Note that even though our
  proof methods fully support parametric specifications, we instantiated 
  some of the parameters for the experiments because the underlying
  provers Yices and ARMC do not support non-linear constraints.}
The table lists execution times for the involved tools: (sys) contains the times
needed by Syspect and the PEA toolkit to write
the TCS, (hpi) the time of \hpilot
to compute the reduction and to check satisfiability with Yices as back-end,
(yic) the time of Yices to check the proof tasks without
reductions by \hpilot.
Due to some
semantics-preserving transformations during the translation process the
resulting TCS consists of 46 transitions. Since our invariant $\sf (Inv)$ is too
complex to be handled by the clausifier of \hpilot, we check the invariant for
every transition in two parts yielding 92 proof obligations.  In
addition, results for the most extensive proof obligation are stated:
one part of the speed update. Further, we performed 
tests to ensure that the specifications are consistent.

The table shows that the time to compute the TCS
is insignificant and that the overall time to verify all transition updates with
Yices and \hpilot does not differ much. On  the
speed update \hpilot was 5 times faster than Yices
alone.
During the development of the case study
\hpilot helped us finding the correct transition invariants by providing models
for satisfiable transitions. 
The table lists our tests with the verification of condition {\sf(Safe)},
which is not inductive over all transitions (cf.\ Sect.~\ref{Sec:Verification}):
here, \hpilot was able to provide a model after 8s  
whereas Yices detected unsatisfiability for 17 problems, returned ``unknown'' 
for 28, and timed out once (listed as (t.o) in the table). 
For the consistency check \hpilot was able to provide a model after 3s, 
whereas Yices answered ``unknown'' (listed as (U)).

In addition, we used ARMC to verify the property ${\sf (Safe_T)}$ of the
timed 
train controller. The full TCS for this proof tasks comprises 8 parallel
components, more than 3300 transitions, and 28 real-valued variables and clocks
(so it is an infinite state system). For this reason, the verification took 26
hours (on a standard desktop computer).



\section{Conclusion}
\label{sec:conclusion}

We augmented existing techniques for the verification of real-time
systems to cope with rich data structures like pointer structures. 
We identified a decidable fragment of the theory 
of pointers, and used it to model systems of trains on linear tracks 
with incoming and outgoing trains. We then proved that certain types of 
complex track systems can be decomposed into 
linear tracks, and that proving safety of train controllers for 
such complex systems can be reduced to proving safety of controllers for 
linear tracks.
We implemented our approach in a new tool chain taking high-level 
specifications in terms of COD as input. 
To uniformly specify processes, data and time, 
\cite{MD1998,AM1998,WC2001} use similar combined 
specification formalisms. We preferred COD due to its strict separation 
of control, data, and time, and
its compositionality (cf.~Sect.~\ref{Sec:ModularSpecifications}),
which is essential for automatic verification. There is also sophisticated
tool support given by \syspect and the PEA toolkit.
Using this tool chain we automatically verified safety properties of a 
complex case study, closing
the gap between a formal high-level
language and the proposed verification method for TCS.
We plan to extend the case study to also consider 
emergency messages (like in
\cite{faber-jacobs-sofronie-07}), possibly coupled 
with updates in the track topology, or updates of priorities. 
Concerning the track topology, 
we are experimenting with more complex axiomatizations  
(e.g.\ for connectedness properties) that are not in the 
pointer fragment presented in Sect.~\ref{Sect:local}; 
we already proved various locality results. 
We also plan to study possibilities of automated invariant generation 
in such parametric systems.  

\smallskip
\noindent {\bf Acknowledgments.} Many thanks to Werner Damm, 
Ernst-R{\"u}diger Olderog and the anonymous referees for their helpful 
comments.






\begin{thebibliography}{10}
\providecommand{\url}[1]{\texttt{#1}}
\providecommand{\urlprefix}{URL }

\bibitem{ADR2009}
Abdulla, P.A., Delzanno, G., Rezine, A.: Approximated parameterized
  verification of infinite-state processes with global conditions. Form.\
  Method Syst.\ Des.  34(2),  126--156 (2009)

\bibitem{AJ1998}
Abdulla, P.A., Jonsson, B.: Verifying networks of timed processes. In: Steffen, B. (ed.) TACAS'98. 
  LNCS, vol. 1384, pp. 298--312. Springer, Heidelberg (1998)

\bibitem{AJN+2004}
Abdulla, P.A., Jonsson, B., Nilsson, M., Saksena, M.: A survey of regular model
  checking. In: Gardner, P., Yoshida, N. (eds.)  CONCUR'04. LNCS, vol. 3170, pp. 35--48. Springer, Heidelberg
  (2004)

\bibitem{AM1998}
Abrial, J.R., Mussat, L.: Introducing dynamic constraints in {B}. In: Bert, D.
  (ed.) B'98. LNCS, vol. 1393, pp. 83--128. Springer, Heidelberg (1998)

\bibitem{APRXZ01}
Arons, T., Pnueli, A., Ruah, S., Xu, J., Zuck, L.D.: Parameterized verification
  with automatically computed inductive assertions. In: Berry, G., Comon, H., Finkel, A. (eds.) CAV'01. LNCS, vol.
  2102, pp. 221--234. Springer, Heidelberg (2001)

\bibitem{manna-vmcai-06}
Bradley, A., Manna, Z., Sipma, H.: What's decidable about arrays? In: Emerson, E.A., Namjoshi, K.S. (eds.) VMCAI'06. 
LNCS, vol. 3855, pp. 427--442. Springer, Heidelberg (2006)

\bibitem{CTV06}
Clarke, E.M., Talupur, M., Veith, H.: Environment abstraction for parameterized
  verification. In: Emerson, E.A., Namjoshi, K.S. (eds.) VMCAI'06. LNCS, vol. 3855, pp. 126--141. Springer,
  Heidelberg (2006)

\bibitem{atr66}
Faber, J., Ihlemann, C., Jacobs, S., Sofronie-Stokkermans, V.: Automatic
  verification of parametric specifications with complex topologies. 
  Reports of SFB/TR 14 AVACS No. 66,
{SFB/TR} 14 {AVACS} (2010), \url{www.avacs.org}

\bibitem{faber-jacobs-sofronie-07}
Faber, J., Jacobs, S., Sofronie-Stokkermans, V.: Verifying {CSP-OZ-DC}
  specifications with complex data types and timing parameters. In: Davies, J., Gibbons, J. (eds.) IFM'07.
  LNCS, vol. 4591, pp. 233--252. Springer, Heidelberg (2007)

\bibitem{HP2007}
Haxthausen, A.E., Peleska, J.: A domain-oriented, model-based approach for
  construction and verification of railway control systems. In: Jones, C.B., Liu, Z., Woodcock, J. (eds.) 
  Formal Methods  and Hybrid Real-Time Systems. LNCS, vol. 4700, pp. 320--348. Springer,
  Heidelberg (2007)

\bibitem{Hoenicke2006}
Hoenicke, J.: Combination of Processes, Data, and Time. Ph.D. thesis,
  {University of Oldenburg}, Germany ({2006})

\bibitem{HO02b}
Hoenicke, J., Olderog, E.R.: {CSP-OZ-DC}: A combination of specification
  techniques for processes, data and time. Nordic J. Comput.  9(4),  301--334
  (2002)

\bibitem{Sofronie-Ihlemann-Jacobs-tacas08}
Ihlemann, C., Jacobs, S., Sofronie-Stokkermans, V.: On local reasoning in
  verification. In: Ramakrishnan, C.R., Rehof, J. (eds.) TACAS'08. LNCS, vol. 4963, pp. 265--281. Springer,
  Heidelberg (2008)

\bibitem{Sofronie-Ihlemann-hpilot08}
Ihlemann, C., Sofronie-Stokkermans, V.: System description: {H-PILoT}. 
  In: Schmidt, R.A. (ed.) CADE'09. LNCS, vol. 5663, pp. 131--139. Springer, Heidelberg (2009)

\bibitem{jacobs-sofronie07}
Jacobs, S., Sofronie-Stokkermans, V.: Applications of hierarchic reasoning in
  the verification of complex systems. ENTCS  174(8),  39--54 (2007)

\bibitem{LahiBrya04}
Lahiri, S.K., Bryant, R.E.: Indexed predicate discovery for unbounded system
  verification. In: Alur, R., Peled, D.A. (eds.) CAV'04. LNCS, vol. 3114, pp. 135--147. Springer,
  Heidelberg (2004)

\bibitem{MD1998}
Mahony, B.P., Dong, J.S.: Blending {Object-Z} and timed {CSP}: An introduction
  to {TCOZ}. In: ICSE'98. pp. 95--104 (1998)

\bibitem{NeculaMcPeak}
McPeak, S., Necula, G.: Data structure specifications via local equality
  axioms. In: Etessami, K., Rajamani, S.K. (eds.) CAV'05. LNCS, vol. 3576, pp. 476--490 (2005)

\bibitem{MFH+2008}
Meyer, R., Faber, J., Hoenicke, J., Rybalchenko, A.: Model checking duration
  calculus: A practical approach. Form.\ Asp.\ Comput.  20(4--5),  481--505
  (2008)

\bibitem{MOR+2008}
M{\"o}ller, M., Olderog, E.R., Rasch, H., Wehrheim, H.: Integrating a formal
  method into a software engineering process with {UML} and {Java}. Form.\
  Asp.\ Comput.  20,  161--204 (2008)

\bibitem{PQ2009}
Platzer, A., Quesel, J.D.: European train control system: A case study in
  formal verification. In: Breitman, K., Cavalcanti, A. (eds.) ICFEM'09. LNCS, vol. 5885, pp. 246--265. Springer,
  Heidelberg (2009)

\bibitem{PR2007}
Podelski, A., Rybalchenko, A.: {ARMC}: The logical choice for software model
  checking with abstraction refinement. In: Hanus, M. (ed.) PADL'07. LNCS,
  vol. 4354, pp. 245--259. Springer, Heidelberg (2007)

\bibitem{Sofronie-cade-05}
Sofronie-Stokkermans, V.: Hierarchic reasoning in local theory extensions. In: Nieuwenhuis, R. (ed.)
  CADE'05. LNCS, vol. 3632, pp. 219--234. Springer, Heidelberg (2005)

\bibitem{Sofronie-getco06}
Sofronie-Stokkermans, V.: Sheaves and geometric logic and applications to
  modular verification of complex systems. ENTCS  230,  161--187 (2009)

\bibitem{sofronie-ijcar10}
Sofronie-Stokkermans, V.: Hierarchical reasoning for the verification of
  parametric systems. In: Giesl, J., H{\"a}hnle, R. (eds.) IJCAR'10. LNAI,
  vol. 6173, pp. 171--187. Springer, Heidelberg (2010)

\bibitem{WC2001}
Woodcock, J.C.P., Cavalcanti, A.L.C.: A concurrent language for refinement. In: Butterfield, A., Strong, G., Pahl, C. (ed.)
  IWFM'01. BCS Elec. Works. Comp. (2001)

\end{thebibliography}
\end{document}